%% file: main.tex
\documentclass{article}
\usepackage{graphicx} 
\usepackage{amsmath}
\usepackage{enumitem}
\usepackage{amssymb}
\usepackage[a4paper, total={6in, 8in}]{geometry}
\usepackage{amsfonts}
\usepackage{tikz} 
\usepackage{hyperref}
\usepackage{algorithm}
\usepackage{amsthm}

\usepackage[super]{nth}
\usepackage[noend]{algpseudocode}
\newtheorem{theorem}{Theorem}
\newtheorem{observation}[theorem]{Observation}

\newtheorem{lemma}[theorem]{Lemma}
\newtheorem{corollary}[theorem]{Corollary}

\theoremstyle{definition}
\newtheorem{definition}{Definition}
\newtheorem*{remark}{Remark}

\DeclareMathOperator{\range}{range}
\newcommand{\none}{\textsc{None}}
\newcommand{\persist}[1]{\(#1\)-\textsc{Persist}}

\usepackage[parfill]{parskip}

\usetikzlibrary{arrows.meta,calc,positioning,decorations.pathreplacing}

\NewDocumentCommand{\interval}{O{0} m m O{} O{black} O{} O{}}{%
  \draw[very thick,#5] (#2,#1) -- (#3,#1);
  \draw[very thick,#5] (#2,#1+0.12) -- (#2,#1-0.12);
  \draw[very thick,#5] (#3,#1+0.12) -- (#3,#1-0.12);
  \IfNoValueF{#4}{\node[above=2pt] at ($(#2,#1)!0.5!(#3,#1)$) {#4};}
  \IfNoValueF{#6}{\node[below] at (#2,#1-0.12) {#6};}
  \IfNoValueF{#7}{\node[below] at (#3,#1-0.12) {#7};}
}

\NewDocumentCommand{\intervalthin}{O{0} m m O{} O{black} O{} O{}}{%
  \draw[#5] (#2,#1) -- (#3,#1);
  \draw[#5] (#2,#1+0.12) -- (#2,#1-0.12);
  \draw[#5] (#3,#1+0.12) -- (#3,#1-0.12);
  \IfNoValueF{#4}{\node[above=2pt] at ($(#2,#1)!0.5!(#3,#1)$) {#4};}
  \IfNoValueF{#6}{\node[below] at (#2,#1-0.12) {#6};}
  \IfNoValueF{#7}{\node[below] at (#3,#1-0.12) {#7};}
}

\makeatletter
\def\BState{\State\hskip-\ALG@thistlm}
\makeatother
\title{Real Time Proportional Throughput Maximization: How much advance notice should you give your scheduler?}
\author{Nadim A. Mottu}
\date{May 2026}

\begin{document}
	\maketitle

	\begin{abstract}
        We will be exploring a generalization of real time scheduling problem sometimes called the real time throughput maximization problem. Our input is a sequence of jobs specified by their release time, deadline and processing time. We assume that jobs are announced before or at their release time. At each time step, the algorithm must decide whether to schedule a job based on the information so far. The goal is to maximize the value of the sum of the processing times of jobs that finish before their deadline, this is often called real time throughput with proportional weights.
    
        We extend this problem by defining a notion of \(t\)-advance-notice, a measure of how far in advance each job is announced relative to their processing time. 
		
		We show that there exists a class of algorithms \persist{\tau} parametrized by some value \(\tau\in [1,\infty)\). If an input sequence has \(t\)-advance-notice, \persist{\tau} is \(\frac{\tau - 1}{\tau^2 +\tau - 1}\)-competitive. In particular, we show that for any \(t \leq \frac{1}{2}\), there is an algorithm that achieves \(\frac{t-t^2}{1+t-t^2}\)-competitiveness and for any \(t \geq \frac{1}{2}\), there is an algorithm that achieves \(\frac{1}{5}\)-competitiveness.
		
		We also give an upper bound of any algorithm that relies on input sequences having \(t\)-advance-notice. We show that the competitive ratio of any algorithm can be at most \(\frac{t}{2t+1}\) against input sequences that have \(t\)-advance-notice. In particular, we show that regardless of how much advance-notice is given, no algorithm can reach $\frac{1}{2}$-competitiveness.
    \end{abstract}

\input{introduction.tex}

    \input{power_of_advanced_notice.tex}

\input{advance-notice-negative.tex}
    \input{conclusion.tex}

	\newpage
    \bibliographystyle{plain}
    \bibliography{ref} 
\end{document}

%% file: introduction.tex
\section{Introduction}   

\subsection{Our model}

Our input is a sequence of jobs $\mathcal{J}=(J_1,\ldots, J_n)$ where $J_i = (a_i, r_i, p_i, d_i) \in \mathbb{R}_{\geq 0}^4$. In this model, each job is announced by an adversary at time $a_i = a(J_i)$, at this time, its release time $r_i=r(J_i)$, processing time $p_i = p(J_i)$ and deadline $d_i = d(J_i)$ are revealed to the algorithm. The algorithm does not know what jobs may be announced in the future or the number of jobs $n$. At each time step, the algorithm $\textsc{Alg}$ must decide whether to schedule a job based on the information so far, subject to the restriction that if a job $J_i$ is scheduled at time $s_i = s(J_i) = s^{\textsc{Alg}}(J_i)$ then no other job can be scheduled during the interval $[s_i, s_i+p_i)$, furthermore, $r_i \leq s_i \leq s_i + p_i \leq d_i$. The goal is to maximize the value of the sum of the processing times $p_i$ of jobs that finish before their deadline, or equivalently the total time any job is running. Later in this paper, we will also briefly explore the alternative goal of maximizing the total value of the ``weights'' of jobs that finish before their deadline. Until we reach this chapter, we assume that the \emph{weight} of a job is equal to its processing time, meaning that maximizing the weights of completed jobs is the same as maximizing the total processing time.

\subsection{Related Works}
The problem of real-time weighted throughput and its applications has been studied quite thoroughly. 

It is not difficult to show that in a deterministic real-time online model, any algorithm can be made to perform arbitrarily bad if no restriction is placed on the input sequence. After all, an adversarial scheduler could announce an arbitrarily small job, then announce another job of arbitrarily large weight overlapping with that job if and only if the algorithm takes the original job. Regardless of what the algorithm does, its performance when compared to an optimal offline algorithm is arbitrarily large. For this reason, most of the existing work related to this problem tackle special cases or restrictions on the problem.

The problem of job scheduling with fixed start and end times, called ``interval'' scheduling, has been studied by Woeginger \cite{WOEGINGER19945} where he found that maximizing the total weights of jobs could be done by a $\frac{1}{4}$-competitive algorithm when the weights are assumed to be given by functions he describes as $C$-benevolent and the algorithm is allowed to ``revoke'' another job, stopping its execution early and forfeiting its value in exchange for scheduling another, better job. Notably, his result applies to the proportional case as well. Later, still focusing on intervals, Fung et al. \cite{Fung2014} gives a $\frac{1}{2}$-competitive ratio barely random algorithm for ``monotone'' intervals as well as $C$-benevolent and $D$-benevolent weight functions. He also gives a $\frac{1}{3}$-competitive algorithm for equal length jobs allowing for variable start times (also known as slack). In \cite{Fung2014}, he assumes that algorithms have the ability to ``restart'' other jobs, which is to say interrupt them similar to ``revoking'', only now we also allow jobs that are revoked to be started again from scratch so long as their deadline has not past. Chrobak et al. \cite{ChrobakRestarts} also studied this model and showed that for equal processing time jobs deterministic algorithms a competitive ratio of $\frac{2}{3}$ is tight when restarting is allowed for unweighted job scheduling. However, they also show that when inputs are given sufficiently large ``slack'', restarting is not needed for their algorithm to reach the desired competitive ratio. The more general case of restarting algorithms for unweighted job scheduling is studied by Hoogeveen et al. \cite{HOOGEVEEN2000193} where he gives a tight $\frac{1}{2}$ competitive ratio for arbitrarily processing times in the unweighted restart model. Changdao He\cite{he2025revokevsrestartunweighted} shows that in contrast to restarting, revoking on its own is not enough to get a constant competitive ratio for job scheduling with unweighted jobs.

On the offline side of the problem, \cite{BaptisteOn4} Baptiste gives a polynomial time and space algorithm for maximizing the number of on time jobs when jobs are allowed to stop and resume. As for the case without resuming, the very closely related task of minimizing weight of late jobs is shown to be NP-Complete by Karp \cite{Karp1972}. A $(1-\frac{1}{e} - \varepsilon )$-approximation is given by Karakostas \cite{KARAKOSTAS2025115258}. Meanwhile, Borodin et al. \cite{Borodin2003} show that there exists a priority algorithm with $\frac{1}{3}$-approximation ratio for unweighted job scheduling and $\frac{1}{4}$-approximation ratio for proportional weights, they also show that no fixed-priority algorithm can achieve better approximation ratio than $\frac{1}{2}$ and $\frac{1}{4}$ for the unweighted and proportional cases respectively.

In general, when studying real-time scheduling, the time of announcement $a_i$ and the time of release $r_i$ are assumed to be the same for all jobs. That is to say, to the best of our knowledge, we are the first to explore a model where jobs may be announced ahead of their release time in a real-time model. The main benefit of this addition is that it allows us to reach a finite, competitive ratio without the need to restart or revoke. The problem of throughput in a real-time context is very well motivated by many applications such as CPU scheduling, among other things. In many contexts, however, it may be inappropriate or detrimental to stop or revoke a job in the middle of its execution. Somebody who booked a room, or a service, may for example find it inconvenient to lose access in the middle of their allotted time. It may, in these circumstances, seem far more reasonable to require users or ``jobs'' to announce their availability and required time in advance. Ultimately, this extension provides what we view as an appropriate alternative to the preemption models studied elsewhere.

\subsection{Preliminary}
Our goal will be to design algorithms to maximize an (asymptotic) competitive ratio defined as follows:
$$\rho(\textsc{Alg})=\lim_{\textsc{Opt}(\mathcal{J})\to\infty}\inf\frac{\textsc{Alg}(\mathcal{J})}{\textsc{Opt}(\mathcal{J})}$$

Where $\textsc{Opt}(\mathcal{J})$ denotes the value of the optimal offline solution on a sequence of jobs $\mathcal{J}=(J_1,\ldots, J_n)$, and $\textsc{Alg}(\mathcal{J})$ is the value of the algorithm on that same set of jobs. 

In general, the inclusion of an announcement time does not necessarily help us against an adversary relative to a model where jobs are announced at their release times. Since announcing a job \(J_i\) such that \(a_i < r_i\) does not impact the value of the optimal offline solution (which knows about every job from the very beginning), an adversary designing the sequence of jobs $\mathcal{J}$ can always make an algorithm perform worse by making $a_i = r_i$ without impacting the optimal solution. We therefore need to place a restriction on the value of $a_i$ to allow our algorithms to be able to reliably use it.

\begin{definition}\label{t-ad}
    A job \(J\) has \(t\)-advance-notice for some constant \(t\in \mathbb{R}_{\geq 0}\) if \((r_i - a_i) \geq t\cdot p_i\). An input sequence has \(t\)-advance-notice if all jobs in it have \(t\)-advance-notice.
\end{definition}
Since jobs are designed by an adversary and giving an earlier $a_i$ only benefits an online algorithm without changing the optimum, we can assume in most instances that $(r_i - a_i) = t\cdot p_i$ (the adversary does not give any more early notice than what is strictly required). In particular, the model without announcement times is the same as a model in which all input sequences have \(0\)-advance-notice. To emphasize this fact, we sometimes say an input sequence or a job \(J_k\) have \emph{strictly} \(t\)-advance-notice if \((r_i - a_i) = t\cdot p_i\) for all \(i\) or for \(i=k\) respectively. We say an input sequence has \emph{no advance notice} if it has strictly \(0\)-advance-notice.
\begin{remark}
    If an input sequence has \(t\)-advance-notice it also has \(t'\)-advance-notice for all \(t' \leq t\).
\end{remark}

%% file: power_of_advanced_notice.tex
\section{The Power of Advance-Notice}

\subsection{An Algorithm for Proportional Weight Throughput with Restarting}
In this section, we will provide a construction for algorithms that achieves \(\frac{t-t^2}{1+t-t^2}\)-competitive ratio on job sequences with \(t\)-advance-notice for \(t \in [0,\frac{1}{2}]\) and \(\frac{1}{5}\)-competitive ratio for \(t \geq \frac{1}{2}\) without preemption.

We define the following algorithm, we call the \(\tau\)-\textsc{Persist} algorithm. For our analysis, it is useful to treat this algorithm as one that can use restarting as a form of preemption. We will then show that when running this algorithm on input sequences that have \(\frac{1}{\tau}\)-advance-notice, this algorithm never needs to interrupt any job.

\begin{algorithm}[H]
    \caption{Code for \(\tau\)-\textsc{Persist} where \(\tau > 1\)}
    \begin{algorithmic}[1]
        \State When a job \(J_k\) is released at time \(t\):
        \State \(J_c \leftarrow\) job currently processed
        \If {\(J_c = \none \)}
        \State Start processing \(J_k\)
        \ElsIf{\(p(J_k) > \tau \cdot J_c\)}
        \State Interrupt \(J_c\)
        \State Start processing \(J_k\)
        \EndIf
        \State When a job \(J_c\) has completed
        \State \(J_k \leftarrow\) the job with the maximum processing time that can be scheduled at time \(t\)
        \State Start processing \(J_k\)
    \end{algorithmic}
\end{algorithm}


In other words, this algorithm processes the largest job it can, but only interrupts a job when it can be replaced with a job with \(\tau\) times more weight. We note that this algorithm uses \emph{restarts}, meaning that if a job is interrupted, it is added to the set of jobs that can be scheduled. This means that it can be scheduled when a job is completed.

Without loss of generality, we assume when analyzing the competitive ratio of the algorithm all jobs are announced at their release times. We will drop this assumption when proving theorem \ref{thm:no_premption_needed}.

We now analyze the performance of \(\tau\)-\textsc{Persist}. This analysis, proceeds very similarly to Woeginger's analysis of the algorithm he calls \(\textsc{Heu}\) \cite{WOEGINGER19945}. The algorithm \(\textsc{Heu}\) acts identically to how \(2\)-\textsc{Persist} acts on input sequences where slack is zero.

We begin with some definitions that will be useful for our charging argument. We define the range of a job \(J_i\) as follows. Suppose that \(J_i\) was scheduled at a point in which the machine was idle and thus did not interrupt another job. Then its \emph{predecessor chain} is an empty list. Otherwise, suppose that \(J_i\) was scheduled interrupting \(J_c\). Then the predecessor chain of \(J_i\) is the predecessor chain of \(J_c\), along with the tuple \((J_c, \sigma'_c)\) appended to the beginning of the list, where \(\sigma'_c\) is the value of start time of \(J_c\) before it was interrupted. The \emph{successor} of \(J_c\) is the highest processing time job that can be scheduled during the time our algorithm is processing \(J_i\). Suppose that the predecessor chain of \(J_i\) is \((\sigma_1', J_1'), \ldots, (\sigma_h', J_h')\) and the successor of \(J_i\) is \(J_c\). We define the \emph{range} of \(J_i\) as the interval \([\sigma_1', \sigma_i + p_i+ p_c]\).

We define the range of a job \(J_i\) as follows: suppose that \(J_i\) was scheduled at a point in which the machine was idle and thus did not interrupt another job. Then its \emph{predecessor chain} is an empty list. Otherwise, suppose that \(J_i\) was scheduled interrupting \(J_c\). Then the predecessor chain of \(J_i\) is the predecessor chain of \(J_c\), along with the tuple \(J_c, \sigma'_c\) appended to the beginning of the list, where \(\sigma'_c\) is the value of start time of \(J_c\) before it was interrupted. The \emph{successor} of \(J_c\) is the highest processing time job that can be scheduled during the time our algorithm is processing \(J_i\). Suppose that the predecessor chain of \(J_i\) is \((\sigma_1', J_1'), \ldots, (\sigma_h', J_h')\) and the successor of \(J_i\) is \(J_c\). We define the \emph{range} of \(J_i\) as the interval \([\sigma_h', \sigma_i + p_i+ p_c]\).

\begin{observation}\label{obs:ranges}
    Consider two ranges \(\range(J_k)\) and \(\range(J_h)\) that are scheduled consecutively by our algorithm. If there is a gap between the two ranges, then any job that can be scheduled in that gap have been completed by our algorithm.
\end{observation}

\begin{proof}
    Suppose for the sake of obtaining a contradiction that there is some job \(J_i\) which can be scheduled at time \(\sigma_i'\) such that \(J_i\) has not been completed elsewhere by the algorithm. Furthermore, suppose that \([\sigma_i', \sigma_i' + p_i)\) intersects the gap between \(\range(J_k)\) and \(\range(J_h)\).
    
    \begin{description}
    \item[Case 1:] \([\sigma_i', \sigma_i' + p_i)\) does not intersect \([\sigma_k, \sigma_k+p_k)\).
    Since the next job that the algorithm schedules is the first job in \(J_h\)'s predecessor chain, no job is being scheduled at \(\sigma_i'\). This means our algorithm should schedule \(J_i\) since it hasn't completed it. This contradicts the fact that \(J_i\) is not in the predecessor chain of \(J_h\).
    \item[Case 2:] \([\sigma_i', \sigma_i' + p_i)\) intersects \([\sigma_k, \sigma_k+p_k)\).
    In this case, \(J_i\) can be scheduled during the time \(J_k\) is being processed and thus has a processing time less than or equal to the processing time of the successor \(J_c\) of \(J_k\). Since \([\sigma_i', \sigma_i' + p_i)\) intersects the gap between \(\range(J_k\) and \(\range(J_h)\). Since \(\sigma_k + p_k + p_c\) is the latest that any job that can be scheduled during \(J_k\) can finish, \(J_i\) cannot intersect the gap. This is a contradiction.
    \end{description}
\end{proof}

\begin{lemma}\label{lem:prop-charging}
    Suppose that we run \(\tau\)-\textsc{Persist} against a \(C\)-benevolent input sequence. Any schedule that is contained in \(\range(J_k)\) which consists only of jobs that have not been completed by our algorithm must have a total weight of less \(\frac{\tau^2}{\tau - 1} \cdot w_k\).
\end{lemma}
\begin{proof}
    Let \((\sigma_1', J_1'),\ldots, (\sigma_h', J_h')\) be the predecessor chain of a job \(J_k\), sorted from earliest to latest scheduled (in other words \(J_h'\) is the earliest job scheduled and \(J_1'\) is the latest) and let \(J_c\) be its successor. We will also let \(p_i' = p(J_i)\). Consider the set of points \(a_{0} = \sigma_k+p_k+p_c, a_{1} = \sigma_k, a_2 = \sigma_k+p_k, a_{3} =\sigma_1',\ldots, a_{h+2} = \sigma_h' \). Their construction can be seen in figure \ref{fig:points}.

\begin{figure}[ht]
    \centering
    \begin{tikzpicture}[x=0.8cm,y=1.2cm,>={Stealth[length=2.5pt]}]

    \interval{0}{15}[\(\range{J_k}\)][black][][]
    \interval[-1]{0}{2}[\(J_2'\)][black][\(\sigma_2'\)][]
    \interval[-5]{0}{15}[][black][\(a_4\)][\(a_0\)]
    \interval[-2]{1}{5}[\(J_1'\)][black][\(\sigma_1'\)][]
    \interval[-5]{1}{15}[][black][\(a_3\)][]
    \interval[-3]{3}{11}[\(J_k\)][black][\(\sigma_k\)][\(\sigma_k +p_k\)]
    \interval[-5]{3}{11}[][black][\(a_2\)][\(a_1\)]
    \interval[-4]{11}{15}[\(J_c\)][black][][\(\sigma_k +p_k+p_c\)]

    \end{tikzpicture}
    \caption{Construction of set of points \(a_0,\ldots,a_{h+2}\)}
    \label{fig:points}
\end{figure}

    By definition of the algorithm, a job is only interrupted when the job that replaces it has \(\tau\) times its processing time. Therefore, \(p_\ell' \leq \tau\cdot p_{\ell+1}' \leq \tau^j \cdot p_{\ell+j}\). This implies that if a job covers the point \(a_j\), its weight is at most \(\tau^{2-j}p_k\) for any \(1 \leq j \leq h\) unless the job was previously completed by the algorithm. For \(j \geq 2\), this follows from the fact that \(a_j\) is the start time of job \(J_{j+2}'\) and thus if a job with weight higher that \(\tau\cdot w_{j+2}\leq \tau^{2-j}w_k\) existed at the time, then it would be scheduled instead. For \(j = 1\) it follows from the fact \(J_k\) was not interrupted and thus no job with weight \(\geq \tau\cdot w_k\) could be scheduled during its execution.

    Consider some schedule \((\sigma_0^*, G_0),\ldots, (\sigma_q^*, G_q)\) consisting of jobs \(G_0,\ldots,G_q\) that the algorithm has not scheduled that is contained in \(\range(J_k)\). Let \(I_0, \ldots, I_q\) be the intervals \[[\sigma_0^*, \sigma_0^* + p(G_0)),\ldots,[\sigma_q^*, \sigma_q^* + p(G_q)).\] We can assume that \(I_0, \ldots, I_q\) cover all of \(\range(J_k)\) since this can only increase the weight of the schedule. We can also assume that every interval \(I_i\) covers at least one of the points \(a_j\) (otherwise we can merge that interval with one of its neighbours and this can once again only increase the weight). This means that the total weight of this schedule is less than \(\sum^\infty_{i=1} \tau^{2-i}w_k = \frac{\tau^2}{\tau - 1} \cdot w_k\), as desired.
\end{proof}

\begin{corollary}
    \(\tau\)-\textsc{Persist} is \(\frac{\tau - 1}{\tau^2 + \tau - 1}\)-competitive.
\end{corollary}
\begin{proof}
    Suppose for the sake of obtaining a contradiction, that there exists some input sequence \(J_1,\ldots, J_n\) such that our algorithm gets the total value \(v\), and there exists a schedule that obtains a value \(v' > \frac{\tau^2+\tau -1}{\tau-1}\cdot v\). Consider this schedule with all the jobs completed by our algorithm removed. By observation \ref{obs:ranges}, all jobs in this schedule have to be scheduled in the range of some job completed by the algorithm. By lemma \ref{lem:prop-charging}, this means that the total value of this schedule \(v'' \leq \frac{\tau^2}{\tau-1}\cdot v\). Since the most value that could have been lost by removing all the jobs completed by the algorithm is \(v\), \(v' \leq v'' + v \leq (\frac{\tau^2}{\tau-1} +1)\cdot v=\frac{\tau^2+\tau -1}{\tau-1}\cdot v\) which is a contradiction.
\end{proof}

\subsection{Augmenting with Advance-Notice}

Our augmented algorithm, which we will call \(\tau\)-\(\textsc{Persist}^*\) works as follows, we simply run \(\tau\)-\textsc{Persist} on the set of known jobs and simulate the algorithm into the future, as if no new jobs get announced. When a job would normally be scheduled by \(\tau\)-\textsc{Persist}, but the algorithm knows (from simulating \(\tau\)-\textsc{Persist} into the future) that it will be preempted by a job yet to be released, the algorithm does not schedule the job, although it still interrupts the currently running job if there is one.

\medskip
\begin{lemma}
    The schedule constructed by \(\tau\)-\(\textsc{Persist}^*\) and by \(\tau\)-\(\textsc{Persist}\) are the same if run on the same input sequence.
\end{lemma}
\begin{proof}
    We fix an input sequence \(\mathcal{J}\). 
    
    By definition, \(\tau\)-\(\textsc{Persist}^*\) only ever schedules a job if \(\tau\)-\(\textsc{Persist}\) does. It also interrupts the execution of a job whenever \(\tau\)-\(\textsc{Persist}\) does. As such it suffices to show that every job completed by \(\tau\)-\(\textsc{Persist}\) is completed by \(\tau\)-\(\textsc{Persist}^*\) at the same time. 
    
    Suppose for the sake of obtaining a contradiction that there are some jobs that are completed by \(\tau\)-\(\textsc{Persist}\) that are not completed by \(\tau\)-\(\textsc{Persist}^*\) at the same time. Let \(J_i\) the earliest such job that is completed by \(\tau\)-\(\textsc{Persist}\).

    Let \(\sigma_i\) be the time at which \(\tau\)-\(\textsc{Persist}\) started processing \(J_i\) when it was completed. Since \(\tau\)-\(\textsc{Persist}^*\) did not schedule this job, by definition, there exists some job \(J_j\) such that \(a_j \leq r_i\), \(w_j > 2\cdot w_i\), and \(r_j \in (\sigma_i, \sigma_i + p_i)\).

    At time \(r_j\), suppose that \(J_i\) is being run by \(\tau\)-\(\textsc{Persist}\). Then by definition of the algorithm, it should interrupt \(J_i\) allow for \(J_j\) to be scheduled. Otherwise, \(J_i\) has since been interrupted by another job, we get a contradiction since \(J_i\) must be completed when scheduled at time \(\sigma_i\).
\end{proof}

By this lemma, we can define the \emph{predecessor chain}, \emph{successor}, and \emph{range} of a job \(J\) being scheduled by \(\tau\)-\(\textsc{Persist}^*\) to be the predecessor chain, successor, and range of this job if it were scheduled by \(\tau\)-\(\textsc{Persist}\). We also define the \emph{predecessor}, or \(J\) to be the first element in the predecessor chain of \(J\). Note that the first element in the predecessor is the last job to be interrupted before \(J\) is scheduled since jobs are appended to the start of the predecessor chain.

\medskip
\begin{theorem}\label{thm:no_premption_needed}
    Suppose that some job \(J_k\), in an input sequence with proportional weights, is scheduled by \(\tau\)-\(\textsc{Persist}^*\) (but possibly not completed) and has \(\frac{1}{\tau}\)-advance-notice. Then the predecessor of \(J_k\) was not interrupted by \(\tau\)-\(\textsc{Persist}^*\).
\end{theorem}

\begin{proof}
    If \(J_k\) has no predecessor, then \(J_k\) did not interrupt a job before being scheduled. Otherwise, let\(J_\ell\) be the predecessor of \(J_k\) and let \(\sigma_\ell\) be the time at which \(J_\ell\) was scheduled by \(\tau\)-\(\textsc{Persist}\) before it was interrupted.

    Since \(J_k\) interrupted \(J_\ell\), \(r_k \in (\sigma_\ell,\sigma_\ell+p_\ell)\). Furthermore, since \(J_k\) has \(\frac{1}{\tau}\)-advance-notice, \((r_k - a_k) \geq \tau^{-1}\cdot p_k\). Since the input sequence has proportional weights, we know that \(p_\ell = w_\ell\) and \(p_k = w_k\). And finally, since \(J_k\) interrupted \(J_\ell\), \(p_\ell \cdot \tau < p_k\). Since \(a_k \leq r_k - \tau^{-1}\cdot p_k < \sigma_\ell + p_\ell - \tau^{-1}\cdot p_k <\sigma_\ell + p_\ell - p_\ell =\sigma_\ell\) we get \((r_k - a_k) \geq \tau^{-1}\cdot p_k\)

    Therefore, since \(a_k < \sigma_\ell\), \(J_k\) was announced before \(J_\ell\) was scheduled, and thus by definition, \(\tau\)-\(\textsc{Persist}^*\) does not schedule \(J_\ell\).
\end{proof}

\begin{corollary}
    For any \(0<t \leq \frac{1}{2}\), there is an algorithm that achieves \(\frac{t-t^2}{1+t-t^2}\)-competitiveness against input sequences with \(t\)-advance-notice and proportional weights without using preemption. Furthermore, for any \(t \geq \frac{1}{2}\), there is an algorithm that achieves \(\frac{1}{5}\)-competitiveness under the same conditions.
\end{corollary}

\medskip
\begin{remark}
    As noted in a previous remark, when slack is zero, we get a slightly better competitive ratio from \(\tau\)-\textsc{Persist}, which becomes \(\frac{\tau -1}{\tau^2}\)-competitive. It follows that this better competitive ratio also carries to the case of advance-notice.
\end{remark}

%% file: advance-notice-negative.tex
\section{The limitations of advance-notice}
\subsection{A Negative Result for Preemption-Free Advance Notice with
proportional weights}
\begin{theorem}
    For every \(t\) and for any throughput maximization algorithm that does not use preemption, there exists an input sequence with proportional weights and \(t\)-advance-notice such that the competitive ratio of the the algorithm against this input sequence is \(\frac{t}{2t+1} - \delta\) for any \(\delta > 0\).
\end{theorem}

\begin{proof}
    We will show that for any $\varepsilon > 0$ and any online algorithm $\textsc{Alg}$ there exists a sequence of jobs $\mathcal{J}$ such that $\textsc{Alg} \leq \textsc{Opt} \cdot \frac{t}{2t+1} + \varepsilon$.

    First we fix $\gamma \leq \frac{\varepsilon (t\cdot (2t+1))}{2+t}$.
    
    First we announce $J_1$ with $a_1 = 0$, $r_1 = t$, $p_1 = 1$ and $d_1$ arbitrarily large.
    
    If $J_1$ is never scheduled by $\textsc{Alg}$ then the adversary declares it to be the last job and we have infinite competitive ratio.
    
    If $J_1$ is scheduled at time $s_1$ then at $s_1 + \gamma$ we announce $J_2$ with $a_2 = s_1+ \gamma$, $r_2 = s_1+ 1 - 2\cdot \gamma$, $p_2 = \frac{p_1 - \gamma}{t}$ and $d_2 = r_2 + p_2$. Note that $J_2$ is impossible to schedule since no preemption is permitted.
    Next the adversary announces $J_3$ with $a_3 = s_1 + \gamma$, $r_3 = s_1+ 2\cdot \gamma$ $p_3 = \frac{\gamma}{t}$ $d_3 = r_3 + p_3$. The adversary will also announce: $J_i$ with $a_i = s_1 + \gamma $, $r_i = d_{i-1}$, $p_i = d_i - r_i$ and $d_i = \min(r_i + \frac{\gamma}{t}, r_2)$. For $i > 3$ until the interval $[s_1, s_1 + p_1]$ is covered.
    
    In this schedule, the only job that the algorithm has done is $J_1$ which has value $p_1$. Meanwhile an optimal algorithm $\textsc{Opt}$ can schedule all jobs since none of them overlap except for $J_1$, and since $J_1$ has arbitrarily large slack, we can always fit it somewhere else.
    
    So the value of $\textsc{Alg} = p_1 = 1$ and $\textsc{Opt} = p_1 + p_2 + p_3 + \sum_i p_i = 1 + \frac{1 - \gamma}{t} + (1 - 2\gamma) = \frac{2t + 1 - 2\gamma - \gamma\cdot t }{t}= \frac{2t + 1}{t} - \frac{2\cdot\gamma + \gamma\cdot t}{t}$.
    
    So $\textsc{Opt} \cdot \frac{t}{2t+1} = \textsc{Alg} - \frac{(2\gamma + \gamma \cdot t)\cdot}{t\cdot (2t+1)}$.  By our choice of $\gamma$ we get that $\textsc{Alg} \leq \textsc{Opt} \cdot \frac{t}{2t+1} + \varepsilon$
\end{proof}
\begin{figure}[ht]
    \centering
    \begin{tikzpicture}[x=0.8cm,y=1.2cm,>={Stealth[length=2.5pt]}]
    \node[draw] at (0,0) {\(\textsc{Alg}\)};
    \interval{2}{5}[][dotted][\(a_1\)][]
    \interval{2}{2}[][black][][]
    \interval{5}{5}[][black][\(r_1\)][]
    \interval{20}{20}[][black][][\(d_1\)]
    \interval{7}{10}[\(J_1\)][blue][][]

    \node[draw] at (0,-1) {\(\textsc{Opt}\)};
    \interval[-1]{2}{5}[][dotted][\(a_1\)][]
    \interval[-1]{2}{2}[][black][][]
    \interval[-1]{5}{5}[][black][\(r_1\)][]
    \interval[-1]{20}{20}[][black][][\(d_1\)]
    \interval[-1]{7}{7}[][black][\(a_2,a_3\)][]
    \interval[-1]{7.2}{7.4}[][olive][][]
    \interval[-1]{7.4}{7.6}[][olive][][]
    \interval[-1]{7.6}{7.8}[][olive][][]
    \interval[-1]{7.8}{8.0}[][olive][][]
    \interval[-1]{8.0}{8.2}[][olive][][]
    \interval[-1]{8.2}{8.4}[][olive][][]
    \interval[-1]{8.4}{8.6}[][olive][][]
    \interval[-1]{8.6}{8.8}[][olive][][]
    \interval[-1]{8.8}{9.0}[][olive][][]
    \interval[-1]{9.0}{9.2}[][olive][][]
    \interval[-1]{9.2}{9.4}[][olive][][]
    \interval[-1]{9.4}{9.6}[][olive][][]
    \interval[-1]{9.6}{9.8}[][olive][][]
    \interval[-1]{9.8}{10}[][olive][][]
    \interval[-1]{7.2}{10}[\(J_3,\ldots,J_n\)][olive][][]
    \interval[-1]{10}{13}[\(J_2\)][red][][]
    
    \interval[-1]{14}{17}[\(J_1\)][blue][][]
    
    \end{tikzpicture}
    \caption{$\textsc{Alg}$ vs $\textsc{Opt}$ schedule}
    \label{fig:advvsopt}
\end{figure}

\medskip
\begin{corollary}
    Regardless of how much advance-notice a sequence of jobs is guaranteed, no algorithm can ever have a competitive ratio of $\frac{1}{2}$ or better. 
\end{corollary}

\subsection{Beyond Proportionality}
\input{advance-notice-c-and-D.tex}

%% file: advance-notice-c-and-D.tex
Restricting ourselves to proportional weights might in some situations seem too strict. We will now look at the performance of algorithms in cases where we are not dealing with proportional weights and show that against certain \(C\)-benevolent and \(D\)-benevolent input sequences, regardless of $t$, an algorithm running on a set of jobs with $t$-advance-notice can be made to perform arbitrarily poorly and cannot achieve a constant competitive ratio.

\medskip
\begin{theorem}
    For any throughput maximization algorithm \(\textsc{Alg}\) that does not use preemption and any $t \in (0,1)$, there exists a C-benevolent function $f$ and a sequence of $f$-related jobs such that any algorithm \(\textsc{Alg}\) can have arbitrarily bad competitive ratio against this sequence of jobs.
\end{theorem}

\begin{proof}
        
    We will show that we can generate a sequence of jobs where \(\textsc{Alg}\)=1 and \(\textsc{Opt}\)=$N$ for any $N$.
    
    We once again use an adversarial argument. Our adversary chooses $f(x) = x^{\log_{\left(\frac{1-\varepsilon}{t}\right)}(N)}$ to be our function $f$ for some very small epsilon. We note that this function is $C$-benevolent for $N$ sufficiently larger than $t$.
    
    Our adversary first announces $J_1$ with $a_1 = 0$, $r_1 = t$, $p_1 = 1$, $d_1 = t+1$ and $w_1 = f(1)$.
    
    If our algorithm does not take the job the adversary declares it to  be the last, we can assume that the algorithm takes the job.
    
    Our adversary next announces $J_2$ with $a_2 = t+\epsilon$, $r_2 = t+1-2\varepsilon$, $p_2 = \frac{1 - \varepsilon}{t}$ $d_2 = r_2 + p_2$ which the algorithm cannot schedule.
    
    The optimal algorithm can take $J_2$ which has value $f(\frac{1 - \varepsilon}{t}) = (\frac{1 - \varepsilon}{t})^{\log_\frac{1-\varepsilon}{t}(N)} = N$ and the algorithm can only take $J_1$ which has value $f(p_1) = f(1) = 1$. We have an arbitrarily large competitive ratio as desired.
\end{proof}

\begin{theorem}
    For any $t$ and any throughput maximization algorithm \(\textsc{Alg}\) that does not use preemption, there is a sequence of unweighted jobs that has $t$-advanced notice such that \(\textsc{Alg}\) does not achieve a finite competitive ratio against this input sequence. In particular, since \(f(p) = 1\) is a D-benevolent function, there exists a D-benevolent function $f$ and a sequence of $f$-related jobs such that any algorithm \(\textsc{Alg}\) can have arbitrarily bad competitive ratio against this sequence of jobs..
\end{theorem}
\begin{proof} 
    Once again we show that we can generate a sequence of jobs where \(\textsc{Alg}=1\) and $\textsc{Opt}=N$ for any $N$.

    Our adversary first announces $J_1$ with $a_1 = 0$, $r_1 = t$, $p_1 = 1$, $d_1 = t+1$ and $w_1 = f(1)$.

    We then announce jobs $J_2,\ldots, J_{N+1}$ such that $\forall i,j \in \{2,\ldots,N+1\}$: $[a_i,d_i] \subset (r_1, d_1)$ and $[a_i,d_i] \cap [a_j,d_j] = \emptyset$. Since all jobs have the same weight, the value of a algorithm that schedules $J_2,\ldots, J_{N+1}$ is $N$, as such \(\textsc{Opt}\) has value at least $N$ but $\textsc{Alg} = 1$ as desired.
\end{proof}

%% file: conclusion.tex
\section{Conclusion, Open Problems and Extensions}

In this paper we investigated and discussed the benefits and limitations of using advance-notice in real-time online scheduling.

We first showed that there exists an algorithm which is $\frac{t-t^2}{1+t-t^2}$-competitive on job sequences that have $t$-advance-notice for $t\in (0,\frac{1}{2}]$ and \(\frac{1}{5}\)-competitive for \(t > \frac{1}{2}\). Therefore, the answer to the question in the title of ``How much advance-notice should you give your scheduler'' is ``\(\frac{1}{2}\)''.

We also showed that no algorithm can achieve better than $\frac{t}{2t+1}$-competitive ratio for any $t$. An open question left by this paper to find a tight bound for preemption-free advance notice. That is to say, can we find an algorithm which is \(\frac{t}{2t+1}\)-competitive? Or is it possible to prove that no algorithm can do better than \(\frac{t-t^2}{1+t-t^2}\)?

In the final section we showed that advance-notice does not help us much with some of the more ``obvious'' weight functions that people consider. As such, an interesting open question would be to see if we could get better results with a different definition of advance-notice that correspond more closely to the weight function used.

\subsection{Acknowledgement}

We are very appreciative of the feedback and encouragement from Allan Borodin, and Changdao He who have provided valuable feedback and discussions throughout the construction of this paper.

\newpage